\documentclass[12pt]{article}
\usepackage{amsmath}
\usepackage{amssymb}
\usepackage{amsthm}
\usepackage{mathrsfs}
\usepackage{epsfig}
\usepackage[usenames]{color}
\usepackage{graphicx}
\renewcommand{\theequation}{\mbox{\arabic{section}.\arabic{equation}}}

\newtheorem{propos}{}[section]
\newcommand{\bprop}{\begin{propos}}
\newcommand{\eprop}{\end{propos}}

\newcounter{Roman}
\addtocounter{Roman}{1}

\newcommand{\beq}{\begin{equation}}
\newcommand{\eeq}{\end{equation}}
\newcommand{\bea}{\begin{eqnarray}}
\newcommand{\eea}{\end{eqnarray}}
\newcounter{saveeqn}

\newcommand{\he}{\hat{e}}
\newcommand{\homega}{\hat{\omega}}

\newcommand{\rd}{\rm d}

\newcommand{\cD}{{\cal D}}
\input epsf
\begin{document}  

\hfill \vbox{\hbox{}} 
\begin{center}{\Large\bf General Relativity as the effective theory of GL(4,R) 
spontaneous symmetry breaking}\\[2cm] 
{\bf E. T. Tomboulis\footnote{\sf e-mail: tombouli@physics.ucla.edu}
}\\
{\em Department of Physics and Astronomy, UCLA, Los Angeles, 
CA 90095-1547}
\end{center}
\vspace{1cm}

\begin{center}{\Large\bf Abstract}
\end{center}  
We assume a $GL(4,R)$ space-time symmetry which is 
spontaneously broken to $SO(3,1)$. 
We carry out the coset construction of the effective theory for the non-linearly realized broken symmetry in terms of the Goldstone fields and matter fields transforming linearly under the unbroken 
Lorentz subgroup. We then identify functions of the Goldstone and matter fields that transform 
linearly also under the broken symmetry. Expressed in terms of these quantities the effective theory 
reproduces the vierbein formalism of General Relativity with general coordinate invariance 
being automatically realized non-linearly over $GL(4,R)$. The coset construction makes no 
assumptions about any underlying theory that might be responsible for the assumed symmetry breaking. We give a brief 
discussion of the possibility of field theories with $GL(4,R)$ rather than Lorentz space-time symmetry providing the underlying dynamics. 

\vfill

\pagebreak

\section{Introduction}
The discovery of broken chiral symmetry in the sixties and the subsequent development of 
chiral Lagrangians apparently first gave rise to  
the idea among particle physicists that General Relativity (GR) might also be an effective theory of the same sort.      
The similarity of the structure of GR to that of effective Lagrangians 
of spontaneously broken symmetries is indeed rather obvious. 
Such effective Lagrangians assume a highly geometrical appearance being field theories on 
homogeneous (coset) spaces, and are formulated in terms of covariant derivatives, curvatures etc. 
In this connection, recall also that in GR both the metric field and its inverse occur in the action, a hallmark of effective Lagrangians. 
An inverse for a  
field quantized as an elementary field makes of course no sense in general,
but inverse fields occurs naturally in 
effective actions for broken symmetries with their exponential parametrizations of Goldstone fields.  

The possibility that this analogy holds for GR is attractive since it obviates 
 quantization of the gravitational field as an elementary field and, as in the case of broken chiral symmetry and QCD, might point to a more fundamental, presumably 
 UV complete, underlying theory responsible for the symmetry breaking. 
 
 The broken symmetry in the case of GR must apparently be a space-time symmetry. Starting with the work of Bjorken, it was indeed suggested that the graviton is the Goldstone boson of spontaneously broken Lorentz invariance \cite{Bj}.   
This has been revived in recent years \cite{KT} - \cite{T}. A gravity theory along these lines, though  
approximating GR at low energies, has to fundamentally differ from GR since it contains actual Lorentz symmetry breaking. 

In this paper we pursue the same idea but consider symmetry breaking leaving Lorentz symmetry intact. 
Specifically, we assume a $GL(4,R)$ space-time symmetry which is spontaneously broken to its 
Lorentz subgroup $SO(3,1)$. We then apply the standard particle physics `coset construction'  \cite{CCWZ} -\cite{V} of the effective theory in which the broken symmetry is realized non-linearly in terms of the resulting Goldstone bosons and matter fields transforming linearly under the unbroken subgroup. In this case there are ten Goldstone bosons transforming like a rank-2 symmetric tensor under $SO(3,1)$. Having obtained the basic elements of the construction, we proceed, as it is customary, to look for functions of the Goldstone and particle fields that have simpler, in particular linear transformation properties under $GL(4,R)$ rather than only the unbroken $SO(3,1)$. Such functions can indeed be straightforwardly obtained. There is, in particular, a distinguished rank-2 $GL(4,R)$ tensor related to the existence of an invariant tensor, the 
Minkowski $\eta_{\mu\nu}$, in the unbroken $SO(3,1)$. The effective theory expressed in terms of these new fields turns out to 
reproduce the standard GR framework, the Hilbert-Einstein term being the leading term in 
the effective action, and with coordinate invariance automatically realized over $GL(4R)$.  

It should perhaps be pointed out that, though some mathematical manipulations are similar, this construction has nothing to do with the many attempts to recast GR as a gauge theory of sorts - in 
this extensive and contorted literature the term `Goldstone boson' is occasionally invoked in various guises. In this paper the point of view is the complete opposite: no fundamental fields or connections gauging some symmetry groups are introduced. Instead, spontaneous breaking (by some 
unknown underlying dynamics) is assumed to have occurred, and the effective theory of the resulting Goldstone degrees of freedom is seen, when expressed in the right variables, to assume the form of gravitational interactions.   
There are in fact only two works \cite{ISS}, \cite{BO} the author is aware of where $GL(4,R)$ spontaneous breaking in connection to gravity as a Goldstone boson is considered in a similar vein. We comment on the relation to
these earlier works in more detail below.

A salient feature of the coset construction \cite{CCWZ} - \cite{V} is that only 
a symmetry breaking pattern is postulated and no assumptions whatsoever are made concerning 
the nature or dynamics of an underlying theory responsible for the symmetry breaking. 
An obvious question then is whether such an underlying theory can be envisioned in the case of 
$GL(4,R)$ spontaneous breaking. Since this symmetry is here taken as a space-time symmetry, 
a natural suggestion would be to consider a field theory model where the Lorentz $SO(3,1)$ group  
is replaced by $GL(4,R)$ (or $SL(4,R)$) as the global symmetry of space-time.  
Such a $GL(4,R)$-invariant theory will be physically very different from Lorentz-invariant theories, the familiar features of the latter, in particular the usual notion of single-particle states, emerging only after symmetry breaking to the Lorentz subgroup. We give 
some discussion in the last section below.

\section{Spontaneously broken $SL(4,R)$ and General Relativity} 
\setcounter{equation}{0}
\setcounter{Roman}{0}
The generators of $GL(4,R)$ can be grouped in the six 
antisymmetric $J^{\alpha\beta}$ generators of its Lorentz subgroup, and the remaining 
10 symmetric linear generators $T^{\alpha\beta}$ comprising nine (shape changing, volume preserving) shear generators and one (volume changing) 
dilatation generator. For brevity, we refer to all linear symmetric transformations as shears. The Lie algebra is given by: 
\bea 
[J^{\alpha\beta}, J^{\gamma\delta}] & = & -i   \left(\, \eta^{\alpha\gamma} J^{\beta\delta} -
\eta^{\alpha\delta} J^{\beta\gamma} - \eta^{\beta\gamma} J^{\alpha\delta} + \eta^{\beta\delta} J^{\alpha\gamma}\,\right)                                               \label{CR1}\\
\ [J^{\alpha\beta}, T^{\gamma\delta}] & = &  -i \left(\, \eta^{\alpha\gamma} T^{\beta\delta} +
\eta^{\alpha\delta} T^{\beta\gamma} -\eta^{\beta\gamma} T^{\alpha\delta} - \eta^{\beta\delta} T^{\alpha\gamma}\,\right)                                             \label{CR2}\\
\ [T^{\alpha\beta}, T^{\gamma\delta}] & = &  i\left(\, \eta^{\alpha\gamma} J^{\beta\delta} +
\eta^{\alpha\delta} J^{\beta\gamma} +\eta^{\beta\gamma} J^{\alpha\delta} + \eta^{\beta\delta} J^{\alpha\gamma}\,\right)  \, .                                           \label{CR3} 
\eea
Adjoining the translations 
\bea
[J^{\alpha\beta}, P^\gamma] & = &   -i (\, \eta^{\alpha\gamma} P^\beta   - 
\eta^{\beta\gamma} P^\alpha \,)                                                              \label{CR4}\\
\  [T^{\alpha\beta}, P^\gamma] & = &   -i (\, \eta^{\alpha\gamma} P^\beta + 
\eta^{\beta\gamma} P^\alpha \,)      \,,                                                     \label{CR5}
\eea
we have the algebra of the Affine group. The Euclidean version of these commutation relations are obtained by the replacement  $\eta^{\alpha\beta} \to -\delta^{\alpha\beta}$. 

\subsection{Coset construction for $GL(4,R)$ breaking}
We now consider the spontaneous breaking of the group  $G=GL(4,R)$ to its subgroup $H=SO(3,1)$. We apply the standard `coset construction' \cite{CCWZ}. Group elements $g\in G$ are 
uniquely decomposed as $g=\gamma h$ where $h\in H$ and 
\beq
\gamma(\xi) =  \exp[\, {i\over 2} \xi\cdot T\,]  \, \in G/H   \;. \label{gdec1}
\eeq
We use the obvious notations $
\xi \cdot T= \xi_{\alpha\beta}T^{\alpha\beta}$, $ u\cdot J = u_{\gamma\delta}J^{\gamma\delta}$.  
This decomposition  amounts to the `canonical' parametrization of the left cosets $G/H$ by means of the 
parameters (`preferred fields') $\xi_{\alpha\beta}$. (Any other parametrization leads of course to equivalent results.) For the action of $g\in G$ on the cosets one can then write 
\beq
g\exp[\,{i\over2} \xi(x)\cdot T\,] = \exp[\,{i\over2} \xi^\prime (x)\cdot T\,] 
\exp [\, {i\over 2}u(\xi(x),g) \cdot J\,] \,.  \label{nonlin1}
\eeq
Also, let $\psi$ denote any field transforming under a linear representation $R(h)$, $h\in H$, of the unbroken subgroup. Then the transformations   
\beq
g:\quad  \xi \to \xi^\prime(\xi,g)\,, \qquad \psi\to \psi^\prime=R(e^{u(\xi,g)\cdot J})\psi \label{nonlin2}
\eeq
give a non-linear realization of $G$. If $g\in H$, the transformations (\ref{nonlin2}) become linear, 
with $\psi$ transforming under $R$ and $\xi$ transforming according to the representation $\cal R$ 
of $H$ determined from 
\beq 
hTh^{-1} = {\cal R}(h) T \,,  \label{gentrans1}
\eeq
i.e., in the present case, as a rank-2 symmetric tensor. 

So far we actually have treated $G$ as an internal symmetry (all equations above hold at  
space-time point $x$). To treat it as a space-time symmetry, 
it is very convenient, though not essential, to use the extension of the coset construction for 
space-time symmetries introduced in \cite{V}. One defines the non-linear realization of $G$ 
as a spacetime symmetry by replacing (\ref{nonlin1}) by 
\beq
g \exp[\,ix^\mu P_\mu\,] \exp [\, {i\over 2}\xi(x)\cdot T\,] = 
\exp[\,ix^{\prime \mu}P_\mu\,] \exp[\,{i\over2} \xi^\prime (x^\prime)\cdot T\,] 
\exp [\, {i\over 2}u(\xi(x),g) \cdot J\,]  \,. \label{nonlin3}
\eeq 
This defines a non-linear realization of $G=GL(4,R)$ on the coordinates and the fields $\xi(x)$. 
Indeed, as it is easily verified by use of (\ref{CR4}) - (\ref{CR5}), for any $g\in G$, (\ref{nonlin3}) implies
\beq
x^\mu \to x^{\prime \mu}= A^\mu_{\ \nu}(g) x^\nu \, ,  \label{nonlin4}
\eeq 
with $gP_\mu g^{-1} = A^\nu_{\ \mu}(g) P_\nu$,  
and 
\beq
g \exp [\, {i\over 2}\xi(x)\cdot T\,] =  \exp[\,{i\over2} \xi^\prime (x^\prime)\cdot T\,] 
\exp [\, {i\over 2}u(\xi(x),g) \cdot J\,]  \,. \label{nonlin5}
\eeq
Here $A(g) = S(q)\Lambda(h)$ denotes the vector representation of $g=qh$, $h\in H$, $q\in G/H$.   
Correspondingly, for fields $\psi$ transforming under a linear representation $R(h)$ of $H$, 
\beq
g:\quad  \psi(x)\to \psi^\prime(x^\prime)=R(e^{u(\xi(x),g)\cdot J})\psi(x)\,. \label{nonlin6}
\eeq
Again, for $g=h\in H$, these non-linear realizations reduce to linear transformations under the 
representations $R(h)$ and ${\cal R}(h)$ for $\psi$ and $\xi$, respectively.\footnote{It should be 
pointed out that (\ref{nonlin3}) may actually be taken to define realizations of the Affine group, since, as one easily verifies, taking 
$g=\exp i\epsilon_\mu P^\mu$ induces  translation shifts of the coordinates with 
$\xi^\prime(x^\prime) = \xi(x)$. In fact, local translations $\epsilon_\mu(x)$ may 
be accommodated, which implies that some form of general coordinate invariance must be lurking 
in the resulting effective theory. But we will not follow this line of argument here. \label{Faffine}}

To construct the effective Lagrangian one needs covariant derivatives conveniently 
obtained \cite{CCWZ}  from the Maurer-Cartan form $\gamma^{-1} {\rm d}\gamma$, or, for space-time symmetries, its modification \cite{V} corresponding to (\ref{nonlin3}). So,  letting 
\beq 
\Gamma = \exp [i x\cdot P] \,\gamma(\xi(x)) = \exp [i x\cdot P] \,\exp [{i\over 2}\xi(x) \cdot T]
\;, \label{G}
\eeq
covariant derivatives are obtained by expanding $\Gamma^{-1} \rd \Gamma$ in the group 
generators: 
\beq 
\Gamma^{-1} \rd \Gamma = i \,\homega^\alpha P_\alpha + {i\over 2} D_{\alpha\beta} T^{\alpha\beta} 
+ {i\over 2} \omega_{\alpha\beta} J^{\alpha\beta} \, .\label{Gexp}
\eeq 
Noting that 
\beq
\Gamma ^{-1}\rd \Gamma = \exp[-{i\over 2}\xi\cdot T]\,(i \rd x\cdot P)\,\exp[ {i\over 2}\xi\cdot T] 
+  \exp[-{i\over 2}\xi\cdot T]\,\rd \exp[ {i\over 2}\xi\cdot T] \, ,  \label{Gexp1}
\eeq
a calculation (Appendix A) gives explicitly:
\bea
\homega ^\alpha & = &  \rd x^\mu e_\mu^{\ \alpha}  \label{eform}\\
D_{\alpha\beta}  & = & -{1\over 2}\{\, e^{-1}, \rd  e \, \}_{\alpha\beta}  \\
\label{xicov1}
\omega_{\alpha\beta}&  = & - {1\over 2} [ \, e^{-1}, \rd e \,]_{\alpha\beta} \, . \label{spincon1}
\eea
Here the symmetric matrix $e$ is defined by 
\beq 
e_{\alpha\beta} \equiv ( e^{-\xi})_{\alpha\beta} = (\exp[\,{i\over 2}\, \xi\cdot T_{\rm v}\,]) _{\alpha\beta} \,, \label{ematrix}
\eeq
where in the second equality the subscript ${\rm v}$ denotes generators in the vector 
representation: 
\bea 
{(J^{\alpha\beta}_{\rm v})}_\gamma^\delta & =&  i\left(\, \delta^\alpha_\gamma \eta^{\beta\delta} - \eta^{\alpha\delta} \delta^\beta_\gamma \,\right) \label{vecrep1}\\
{(T^{\alpha\beta}_{\rm v})}_\gamma^\delta & = &  i\left(\, \delta^\alpha_\gamma \eta^{\beta\delta} + \eta^{\alpha\delta} \delta^\beta_\gamma \,\right)  \,. \label{vecrep2}
\eea
The one-forms $\homega^\alpha$ and $D_{\alpha\beta}$ transform covariantly, i.e like the fields $\psi$, whereas $\omega_{\alpha\beta}$ transform inhomogeneously (`like a gauge field'):
\beq
D_{\alpha\beta}=\homega^\gamma {\cal D}_{\gamma\alpha\beta}
\,, \qquad  
\omega_{\alpha\beta} = \homega^\gamma \omega_{\gamma\alpha\beta}  \; .
\label{xicov2} 
\eeq 
$D_{\alpha\beta}$ gives the Goldstone field `covariant derivative' ${\cal D}$.  
Thus, one obtains  
\beq 
{\cD}_{\gamma\alpha\beta} = -{1\over 2}{ e^{-1}_{\ \ \gamma}}^ \kappa 
\{\, e^{-1}, \partial_\kappa e \, \}_{\alpha\beta} \, . 
\label{xicov3}
\eeq
Similarly, $\omega_{\gamma\alpha\beta}$ gives the `spin connections'  and 
serves to define the covariant derivative $D\psi$ of any field $\psi$ transforming as in 
(\ref{nonlin6}):
\beq 
D_\gamma \psi(x)= {e^{-1}_{\ \ \gamma}}^\kappa\partial_\kappa \psi + {i\over 2}\omega_{\gamma\alpha\beta} J^{\alpha\beta}\psi  \,, \label{psicov1}
\eeq
with the generators in the representation of $\psi$. 

As it is well known, however, this covariant derivative is not unique. (\ref{spincon1}), obtained through (\ref{Gexp}), ensures the right transformation properties for $D\psi$. But adding to 
the expression read off (\ref{spincon1}) and (\ref{xicov2}) appropriate covariently transforming terms, such as appropriate (linear) functions of $\cD$, , results into equally good definitions of the covariant derivative of $\psi$. Different such choices 
simply amount to reshuffling of terms in the effective Lagrangian formed from $\cD$, $\psi$ and 
$D\psi$, and are a matter of convenience; a particular choice may result into a more 
transparent structure of the effective Lagrangian.\footnote{ An example is provided 
by the pion-nucleon chiral Lagrangian (in the parametrization choice historically first used
for the pion field), where 
a straightforward way of getting a nucleon covariant derivative gives  a form which obscures the fact that the axial coupling constant, as it occurs in the Goldberger -Treiman relation, is an independent coupling. To avoid this it is preferable to shift a term from this covariant derivative to an interaction with an independent coupling.} 
 For later reference we note, in particular, the  expression 
\bea
\omega_{\gamma\alpha\beta} & = &  -{1\over 2}{ e^{-1}_{\ \ \gamma}}^ \kappa 
[\, e^{-1}, \partial_\kappa e \, ]_{\alpha\beta}  -\cD_{\alpha\beta\gamma} + \cD_{\beta\alpha\gamma}    \nonumber \\
& =& -{1\over 2}{ e^{-1}_{\ \ \gamma}}^ \kappa 
[\, e^{-1}, \partial_\kappa e \, ]_{\alpha\beta}  +{1\over 2}{ e^{-1}_{\ \ \alpha}}^ \kappa 
\{\, e^{-1}, \partial_\kappa e \, \}_{\beta\gamma}  - {1\over 2}{ e^{-1}_{\ \ \beta}}^ \kappa 
\{\, e^{-1}, \partial_\kappa e \, \}_{\alpha\gamma} \quad  \label{spincon2}
\eea
which augments the expression following from (\ref{spincon1}) by the addition of the second and third terms in (\ref{spincon2}). 
Note that (\ref{xicov3}) and (\ref{psicov1}), (\ref{spincon2}),  are 
given, in accordance with (\ref{xicov2}), in the basis provided by the $\homega^\alpha$'s.

Any Lagrangian that is invariant under $H$ and is constructed from $\cD$, $\psi$ and $D\psi$ 
will now be invariant under the full group $G$. The most general effective Lagrangian 
describing the interactions of the Goldstone bosons and the particle fields at scales below the symmetry breaking scale is then given by the most general $H$-invariant function of 
$\psi$, $D\psi$, $\cD$, and their higher covariant derivatives such as $DD\psi$ and $D \cD$.  

It is often the case that a simpler or more transparent form of the effective Lagrangian is 
obtained by actually not using the above canonical parametrization for the Goldstone fields $\xi$, but instead introducing functions of $\xi$ and $\psi$ with simpler transformation 
properties under $G$.\footnote{Again, the chiral baryon-meson Lagrangians provide a well-known example: the common textbook meson (Goldstone) field parametrization by a unitary unimodular matrix transforming as a $(\bar{3},3)$ representation of 
$SU(3)\times SU(3)$ is not the canonical one; but a linearly transforming one, albeit subject to the 
unitary-unimodular constraints, that is much more convenient to work with, and somewhat akin to (\ref{metric}) below. } 
In particular, one may find appropriate functions that transform linearly 
under the full group $G$ rather than only the unbroken group $H$. This is discussed 
in detail in \cite{CCWZ}, where a characterization of such functions is given; here we apply 
the results in \cite{CCWZ} in the present context. 

Take for $\psi$ a covector under $H=SO(3,1)$ with components $v_\alpha$ and let 
\beq
V_\mu = e_\mu^{\ \alpha} v_\alpha \,. \label{Gcovect}
\eeq
Then, as can be seen from (\ref{nonlin3}) - (\ref{nonlin6}), the quantities (\ref{Gcovect}) transform linearly as a covariant vector under $G=GL(4,R)$. Indeed:  
\[
V^\prime(x^\prime) =  (e^{-\xi^\prime(x^\prime)}) v^\prime(x^\prime) = (e^{-\xi^\prime(x^\prime)}) \Lambda(u(\xi(x),g)) v(x)= A(g) (e^{-\xi(x)}) v(x)=A(g) V(x)\, .      
\]
Similarly, it is seen that, given a vector under $H$ with components $v^\alpha$, the quantities $V^\mu = {e^{-1}}^\mu_{\ \alpha} v^\alpha$ transform linearly as the contravariant components of a vector under $G$. More generally, from any Lorentz tensor with components $\psi^{\alpha_1\cdots \alpha_k}_{\beta_1\cdots \beta_l}$, one may obtain an $GL(4,R)$ tensor with components given by:
\beq 
\Psi^{\mu_1\cdots \mu_k}_{\nu_1\cdots \nu_l}
= e_{\nu_1}^{\ \beta_1} \cdots e_{\nu_l}^{\ \beta_l}\, {e^{-1}}^{\mu_1}_{\ \alpha_1} \cdots {e^{-1}}^{\mu_k}_{\ \alpha_k} \psi^{\alpha_1\cdots \alpha_k}_{\beta_1\cdots \beta_l} \,.\label{Gtensor}
\eeq
For fields $\psi$ in spinor representations of the Lorentz group, no corresponding quantities 
transforming linearly under $GL(4,R)$ can be constructed, since there are no finite-dimensional 
spinor representations of $GL(4,R)$.  

Of particular importance is the case when  (\ref{Gtensor}) is applied to the invariant tensor 
$\eta_{\alpha\beta}$ of the Lorentz group. Defining 
\beq 
g_{\mu\nu} \equiv e_\mu^{\ \alpha} e_\nu^{\ \beta} \eta_{\alpha\beta} \, ,\label{metric}
\eeq   
one obtains a $GL(4,R)$ symmetric rank-2 tensor: 
\bea
g: \quad g_{\mu\nu}(x) \to g^\prime_{\mu\nu}(x^\prime) & = &  (e^{-\xi^\prime(x^\prime)})_\mu^{\ \alpha} (e^{-\xi^\prime(x^\prime)})_\nu^{\ \beta} \eta_{\alpha\beta}  \nonumber \\
& = & (e^{-\xi^\prime(x^\prime)})_\mu^{\ \alpha} (e^{-\xi^\prime(x^\prime)})_\nu^{\ \beta}  
\Lambda_\alpha^{\ \gamma}(u(\xi,g))\Lambda_\beta^{\ \delta}(u(\xi,g)) \eta_{\gamma\delta} 
\nonumber \\
&= & A_\mu^{\ \kappa}(g) A_\nu^{\ \lambda}(g) g_{\kappa\lambda}(x)   \,.  
\eea
Its inverse $g^{\mu\nu}$, defined by $g^{\mu\kappa}g_{\kappa\nu}= \delta^\mu_\nu$, 
then also transforms linearly, and is, equivalently, given by $g^{\mu\nu} = {e^{-1}}^\mu_{\ \alpha}  
{e^{-1}}^\nu_{\ \beta} \eta^{\alpha\beta}$.

\subsection{General Relativity as the effective theory}
We now note that the passage to the linearly transforming quantities (\ref{Gtensor}) 
has the form of passage from the (non-holonomic) frame provided by the basis one-forms 
$\homega^\alpha$ and their dual basis vectors $\he_\alpha$ - defined through 
$<\homega^\alpha, \he_\beta>=\delta^\alpha_\beta$  - to a coordinate basis $\{\rd^\alpha, \partial_\alpha\}$.  
The relation between these frames being fixed by (\ref{eform}), i.e. 
\beq 
\homega^\alpha = e_\kappa^{\ \alpha} \rd x^\kappa\;, \qquad  \he_\alpha = {e^{-1}}_\alpha^{\ \kappa} \,\partial_\kappa  \, ,\label{framerel}
\eeq 
the relation between tensor components is then indeed that given by (\ref{Gtensor}). 
Taking this geometric point of view, consider the covariant derivative 
$\nabla$ acting on, say, a vector ${\bf v}= v^\alpha \he_\alpha= V^\mu \partial_\mu$: 
\bea 
\nabla_{\he_\gamma} {\bf v} = \nabla_{\he_\gamma} v^\alpha \he_\alpha &=& \he_\gamma(v^\alpha) \,\he_\alpha + v^\alpha {\omega_{\gamma\alpha}}^\delta \,\he_\delta   \nonumber \\ 
& = & \left[ \, {e^{-1}}_\gamma^{\ \kappa} v^\alpha_{\ ,\kappa} + {\omega_{\gamma\beta}}^\alpha 
v^\beta \,\right] \he_\alpha  \label{covder1} 
\eea
with the standard definition of connection coefficients 
\beq 
\nabla_{\he_\gamma} \he_\alpha = {\omega_{\gamma\alpha}}^\beta \,\he_\beta  \,,\label{condef}
\eeq
and commas denoting, as usual, partial derivatives w.r.t. $x^\kappa$. Taking  the generators $J^{\alpha\beta}$ in (\ref{psicov1}) in the vector representation (\ref{vecrep1}) and comparing to (\ref{covder1}) one then gets 
\beq
\nabla_{\he_\gamma} {\bf v} = (D_\gamma v^\alpha) \,\he_\alpha \, . \label{psicov2}
\eeq 
Specification 
of the connection coefficients ${\omega_{\gamma\alpha}}^\beta $ defines a choice of a particular connection structure. Here they are specified by (\ref{spincon1}), or any other 
equivalent choice in the sense described above
- note that these satisfy  $\omega_{\gamma\alpha\beta}= - \omega_{\gamma\beta\alpha}$.  
On the other hand, in terms of covariant derivative components in the coordinate frame one has  
\bea 
\nabla_{\he_\gamma} {\bf v} & = & \nabla_{{e^{-1}}_\gamma^{\ \beta}\partial_\beta} V^\mu \partial_\mu  \nonumber \\
& = & {e^{-1}}_\gamma^{\ \beta} V^\mu_{\ \ ,\beta}\, \partial_\mu + V^\mu {e^{-1}}_\gamma^{\ \beta} \Gamma^\delta_{\ \beta\mu}\, \partial_\delta   \nonumber \\
& = & {e^{-1}}_\gamma^{\ \beta} \left[\, V^\mu_{\ \ ,\beta} +  \Gamma^\mu_{\ \beta\kappa}V^\kappa \,\right] \partial_\mu  \;  \label{covder2}\\
& = & {e^{-1}}_\gamma^{\ \beta} \,\nabla_\beta V^\mu \, \partial_\mu \label{covder3}
\eea
with the standard notations 
\beq
\nabla_{\partial_\kappa} \partial_\lambda = \Gamma^\mu_{\ \kappa\lambda} \partial_\mu 
\label{Ccon} 
\eeq
for the coordinate frame connection components,  and $\nabla_{\partial_\gamma}\equiv \nabla_\gamma$ for the covariant derivative operator along a coordinate basis vector direction.   
From (\ref{psicov2}) and (\ref{covder3}) we then get the relation between the covariant derivatives 
in the two frames: 
\beq
\nabla_\lambda V^\mu = e_\lambda^{\ \gamma} {e^{-1}}^\mu_{\ \alpha} D_\gamma v^\alpha \,.
\label{covderrel1}
\eeq 
Similarly,  
for any type of tensor one obtains  
\beq 
\nabla_\lambda \Psi^{\mu_1\cdots \mu_k}_{\nu_1\cdots \nu_l}
= e_\lambda^{\ \gamma}   e_{\nu_1}^{\ \beta_1} \cdots e_{\nu_l}^{\ \beta_l}\, {e^{-1}}^{\mu_1}_{\ \alpha_1} \cdots {e^{-1}}^{\mu_k}_{\ \alpha_k} D_\gamma \psi^{\alpha_1\cdots \alpha_k}_{\beta_1\cdots \beta_l} \,,\label{covderrel2}
\eeq
and the obvious extension involving any number of derivatives. 
Note that (\ref{covderrel2}) is consistent with 
(\ref{Gtensor}) as it should, since, by construction, covariant derivatives $D\psi$ of tensors $\psi$ transform as tensors. Furthermore, comparing (\ref{covder1}) and (\ref{covder2}) (or,  
equivalently, combining (\ref{condef}), (\ref{Ccon}) and (\ref{framerel})) gives the relation between the connection components in the two frames: 
\beq 
\Gamma^\mu_{\ \lambda\kappa} = (e^{-1} e_{, \lambda})^\mu_{\ \kappa} - e_\lambda^{\ \gamma} \, {e^{-1}}^{\mu\alpha} \omega_{\gamma\alpha\beta}\, e^\beta_{\ \kappa} \,.
\label{C-spincon}
\eeq

Applying (\ref{covderrel2}) now to the tensor $g_{\mu\nu}$ defined in (\ref{metric}) 
gives
\beq 
\nabla_\lambda g_{\mu\nu} =  e_\lambda^{\ \gamma}  e_\mu^{\ \alpha} e_\nu^{\ \beta}
D_\gamma \eta_{\alpha\beta} = 0  \label{0metricder}
\eeq
since, from (\ref{psicov1}), $D_\gamma\eta_{\alpha\beta}=0$.  

Solving (\ref{0metricder}) for the $\Gamma^\mu_{\ \kappa\lambda}$'s in the familiar way 
gives then 
\beq 
\Gamma^\mu_{\ \lambda\kappa} = {1\over 2} g^{\mu\sigma}\,\left( \, g_{\sigma\kappa, \lambda} + 
g_{\sigma\lambda,\kappa} - g_{\kappa\lambda, \sigma} \,\right)  \label{Ccon1} \, ,
\eeq
i.e. the Christoffel symbols. This in turn determines a set of spin connection coefficients 
through (\ref{C-spincon}). Indeed, inserting (\ref{Ccon1}) into (\ref{C-spincon}) 
determines $\omega_{\gamma\alpha\beta}$ to be given by (\ref{spincon2}).

Defining a curvature tensor in the usual way, i.e. 
\beq 
r^\alpha_{\ \beta\delta\gamma} \, v^\beta \he_\alpha = \left[ \left( \nabla_{\he_\delta} \nabla_{\he_\gamma} - \nabla_{\he_\gamma} \nabla_{\he_\delta}\right)  - \nabla_{[\he_\delta, \he_\gamma]}\right] {\bf 
v}  \, ,  \label{curv1} 
\eeq
a straightforward computation gives  
\beq
\left( \nabla_{\he_\delta} \nabla_{\he_\gamma} - \nabla_{\he_\gamma} \nabla_{\he_\delta}\right) {\bf v} - \nabla_{[\he_\delta, \he_\gamma]} {\bf 
v} = {e^{-1}}_\delta^{\ \epsilon} {e^{-1}}_\gamma^{\ \beta} \,\Big[ ( \nabla_\epsilon\nabla_\beta - \nabla_\beta\nabla_\epsilon ) V^\mu \Big] \partial_\mu  \,.\label{curvrel1}
\eeq
Hence, with 
\beq
R^\mu_{\ \alpha\beta\gamma} V^\alpha =  ( \nabla_\beta\nabla_\gamma - 
\nabla_\gamma\nabla_\beta ) V^\mu  \label{curv2}
\eeq 
defining the curvature tensor components in the coordinate frame, one has the relation 
\beq
{e^{-1}}^\mu_{\ \alpha} e_\delta^{\ \epsilon}e_\gamma^{\ \zeta} e_\kappa^{\ \beta} \, 
r^\alpha_{\ \beta\epsilon\zeta} = R^\mu_{\ \kappa\delta\gamma}  \,, \label{curbrel2}
\eeq 
as indeed required by (\ref{Gtensor}). Note that the last term on the r.h.s. in (\ref{curv1}) is crucial 
for consistency in the non-holonomic frame, whereas, by (\ref{curv2}), $R^\mu_{\ \delta\gamma\kappa}$ is given 
by the familiar Riemann tensor expression in terms of the $\Gamma^\mu_{\ \lambda\kappa}$'s.   

To recapitulate, we carried through the standard coset construction of the spontaneously breaking 
of $G=GL(4,R)$, as a space-time symmetry, to its Lorentz subgroup $H=SO(3,1)$  in terms of massless Goldstone fields and 
matter fields. These provide a non-linear realization of $G$ and a linear 
realization of $H$. We then proceeded to look for functions 
of these fields with linear transformation properties also under $G$. 
Having obtained such quantities we found that their relation to the original fields, as given by (\ref{Gtensor}), 
(\ref{covderrel2}), is that of the transition from a non-holonomic frame to a 
coordinate frame, the transformation being determined by the Goldstone fields. 
Furthermore, the existence of the invariant tensor $\eta_{\alpha\beta}$ in $SO(3,1)$ 
translates into the existence of rank-2 tensor transforming linearly under $GL(4,R)$ whose 
covariant derivative vanishes. This in turn, selects a unique connection, which 
in the coordinate frame description is given by the Christoffel symbols. We noted that  
this transition to (finite-dimensional) fields linearly transforming under $GL(4,R)$ is possible only for tensor, but not spinor, representations, of $SO(3,1)$.

It is clear that we have recovered the basic elements of the General Relativity formalism with 
$g_{\mu\nu}$, defined in (\ref{metric}), serving as the metric tensor. To make this explicit we 
introduce some further notational conventions. 
All quantities introduced above have been defined as functions of the $SO(3,1)$ tensor 
fields $\psi$ and $\xi$ with all indices raised and lowered by $\eta_{\alpha\beta}$.  
In particular, $GL(4,R)$ tensors have been defined as such functions. 
We now write 
\beq E_\mu^{\ \alpha} = e_\mu^{\ \alpha}\, , \qquad E^\mu_{\ \alpha}={e^{-1}}^\mu_{\ \alpha}  \, , \label{vierbein1}
\eeq
and agree to raise and lower indices from the middle of the greek alphabet by $g_{\mu\nu}$, and 
indices from the beginning of the greek alphabet by $\eta_{\alpha\beta}$.  
This is easily seen to be a consistent convention since, as it is easily verified, one has  
\beq
E^{\mu\alpha}E_{\nu\alpha} = \delta^\mu_\nu\,, \quad 
E^{\mu\alpha}E_{\mu\beta} = \delta^\alpha_\beta\,, \quad 
E_\mu^{\ \alpha}E_{\nu\alpha} = g_{\mu\nu}\, , \quad 
E^\mu_{\ \alpha}E^{\nu\alpha} = g^{\mu\nu}\,, \quad 
E_\mu^{\ \alpha}E^{\mu\beta}= \eta^{\alpha\beta} \, ,  \label{vierbein2}
\eeq
and relations such as, for example,  
\[ g_{\mu\kappa}V^\kappa = V_\mu\,, \quad  g^{\mu\kappa}V_\kappa = V^\mu\,, \quad 
{\Psi^\mu}_{\nu\kappa}g^{\kappa\lambda}= {\Psi^{\mu}_{\ \nu}}^\lambda\, , \quad 
{\psi^\alpha}_{\beta\gamma} \eta^{\gamma\delta}= {\psi^{\alpha}_{\ \beta}}^\delta 
=E_\mu^{\ \alpha}E^\nu_\beta E_\lambda^\delta {\Psi^\mu_{\ \nu}}^\lambda     \]
are equivalent in content to (\ref{Gcovect}), (\ref{Gtensor}).

With these conventions then, $E_\mu^\alpha$ serve as a symmetric tetrad (vierbein) 
connecting a local orthonormal frame to a `world' coordinate system with metric 
$g_{\mu\nu}$.  The general effective Lagrangian in terms of these fields, rather than the 
original non-linearly transforming $\xi$'s, is now given by the sum over 
all possible $G$-invariant 
functions of the metric, $\Psi$ and $\nabla \Psi$,  the only possible leading term being 
the Hilbert-Einstein action. Spinor fields couple in the effective action through the vierbein in the usual manner. 

One may ask how the general coordinate invariance present in this effective action came about since such invariance was not input in the original setup;  only the 
spontaneous breaking of $G=GL(4,R)$ was postulated at the outset.  The answer\footnote{In this connection see also footnote \ref{Faffine}.} is implicit in 
the fact that $G$ is treated as a space-time symmetry. 
After going over to variables giving a linear realization of $G$, this means 
that any tensor obtained by covariant differentiation from another tensor, such as 
$\nabla_\lambda \Psi^{\mu_1\cdots \mu_k}_{\nu_1\cdots \nu_l}$, transforms linearly (homogeneously) as a $G$-tensor also w.r.t. space-time indices, such as  $\lambda$, introduced by the differentiation. Any $G$-invariant monomial then remains invariant when the tensors in it are subjected 
to transformations by matrices $A(g)$, $g\in G$ which have been made space-time dependent, in 
particular ${A^\mu}_\nu = \partial x^{\prime\,\mu}/\partial x^\nu$, for any differentiable $x^\prime(x)$. 
In this manner one effectively ends up with a non-linear realization of general coordinate transformations over  
their linearly realized $GL(4,R)$ subgroup, which is indeed how they appear in  the GR formalism.

In summary, the argument is based solely on: 
(i) the assumption that there is a space-time $GL(4,R)$ symmetry which is spontaneously broken to $SO(3,1)$;
(ii) the fact that  $SO(3,1)$ possesses an invariant constant tensor, i.e. $\eta_{\alpha\beta}$. 
With no other assumptions or inputs, application of the coset formalism of spontaneously broken symmetries then leads, by straightforward derivation, to the conclusion above.

\subsection{Relation to previous work}

The relation of the formalism of GR to spontaneous breaking of 
$GL(4,R)$ was previously considered in \cite{ISS} and \cite{BO}. 
 
The authors of \cite{ISS} note the similarity of the formalism of GR to that of non-linear 
realizations of symmetries, and identify $GL(4,R)$ as the natural relevant group. 
For somewhat obscure reasons, however, they choose not to follow through with the complete coset construction for $GL(4,R)$, 
and as a result manage only to suggest rather than arrive at an exact correspondence. 

The authors of \cite{BO} take two groups, the conformal group and $GL(4,R)$ (actually the affine group) as their 
symmetry groups. They first carry out separate coset constructions for each group.  
They then impose what they call 
the  `simultaneous realization' of the two groups by demanding that the $GL(4,R)$ covariant derivative is expressed solely in terms of the conformal group covariant derivative.  
They then argue that this constraint uniquely leads to GR. This procedure, however, is clearly problematic. 
Such a constraint does not appear possible or even meaningful within the usual framework 
of Lagrangian field theory. In any event, as we saw, no such externally imposed constraints are necessary to relate the effective theory of broken $GL(4,R)$ to GR. 
Both these papers, however, make the crucial observation of the existence of linearly transforming quantities such as (\ref{metric}), and their central role in establishing a correspondence with  GR.

\section{Discussion} 
\setcounter{equation}{0}
\setcounter{Roman}{0}
Our starting point was the assumption that there is a $GL(4,R)$ space-time symmetry which 
is spontaneously broken down to $SO(3,1)$. 
The explicit form of the effective action describing physics below the symmetry breaking scale depends on the choice of field parametrizations and covariant derivatives. Different choices lead to different forms of the action which must be equivalent in the sense that they give the same physical amplitudes. 
But this may not always be easy to demonstrate explicitly as it may involve very non-trivial resummations of interactions. Starting from the `canonical' parametrization it is common to seek functions of the Goldstone and matter fields 
that transform linearly also under the broken rather than only the unbroken group. 
In the case of broken $GL(4,R)$ we found that, expressed in terms of such linearly transforming fields, the effective theory assumes the form of gravitational theory in the GR framework, the Hilbert action being the simplest term in the general effective action. Needless to say, this form would be quite obscured in the original canonical or other 
non-linear parametrizations. In particular, the decoupling of Goldstone bosons not corresponding to the physical graviton degrees of freedom would generally not be manifest; and arriving at the equivalence by direct computation of physical amplitudes would be rather nontrivial. 

By design, the construction of the effective theory of broken symmetries \cite{CCWZ} - \cite{V}
makes no assumptions about the nature of any underlying theory that may be responsible for the 
assumed symmetry breaking pattern.
The obvious question then is whether there is an underlying theory with good UV behavior, 
whose dynamics drives $GL(4,R)$ symmetry breaking leading to the effective theory description of its broken $GL(4,R)$ phase given above.

Since here we are concerned with space-time symmetries, the most straightforward and perhaps natural way to proceed is to consider field theories with $GL(4,R)$, or $SL(4,R)$  
replacing $SO(3,1)$ as the global space-time symmetry.\footnote{Another possibility involving alignment of internal and external groups via condensate formation was suggested in 
\cite{T}.}
We will not examine the ingredients necessary for constructing field theories with 
space-time symmetry group $GL(4,R)$, 
or more properly its universal covering group $\overline{GL(4,R)}$ (we will not bother to 
always make  this distinction for the purposes of this discussion), in any detail here. We will only point out a couple of apparently generic features of such field theories. 

First, note that since there is no invariant constant tensor in $GL(4,R)$, i.e. no analog to the 
invariant $\eta_{\alpha\beta}$ of the $SO(3,1)$-symmetric case, there is no natural metric. 
In this sense,  $GL(4,R)$-symmetric space-time is less `rigid' than Minkowski space-time. 
A metric could be introduced only as one of the dynamical fields. Requiring good UV behavior, 
together with this absence of an invariant metric, provides very strong constraints. 
Suppose one looks for the analog of a 
gauge theory of fermion and vector fields. One may, in particular, look for the $\overline{GL(4,R)}$ generalization of the Dirac equation. Taking the fermion field $\psi(x)$ to transform according to an infinite-dimensional representation\footnote{The theory of representations of non-compact groups such as $GL(4,R)$ or $SL(4,R)$ is rather more 
involved and richer, but also less developed, than the familiar representation theory of compact groups. Large classes of representations, including the so-called principal and discrete series,  can be obtained by (one of the variants of) the method of induced representations \cite{K}. 
There are no finite-dimensional spinor representations; and all unitary representations 
are of course infinite-dimensional. Certain representation classes have been studied in 
\cite{Reps}. We note here that the so-called 
multiplicity-free representations (cf. first paper in \cite{Reps}), i.e. those that upon reduction  
to the $SO(3,1)$ subgroup contain representations of the latter only once, are not sufficient for building $\overline{GL(4,R)}$ or $\overline{SL(4,R)}$ invariant field theories. In particular, to construct vector operators defined in (\ref{vector}) non-multiplicity-free representations are necessary.}
$S^{(\nu)}$ (the representation label $\nu$ typically comprises four complex numbers), one needs vector operators $X^\mu$ (analogs 
of the $\gamma^\mu$ matrices) defined by 
\beq 
S^{({\nu}^\prime)}(g^{-1}) X^\mu S^{(\nu)}(g) = {a(g)^\mu}_\nu X^\nu \, ,  \label{vector}
\eeq
where $a(g)$ denotes the fundamental vector representation, $g\in GL(4,R)$. A Dirac operator $X^\mu \partial_\mu$ can then be constructed. In addition, to obtain an invariant Lagrangian, one needs an intertwining operator $\beta$ (analog of $\gamma^0$) between the 
representations $\nu$ and $\nu^{\,\prime}$ connected by the vertex operators. Given appropriate choice of representations, such operators  can be constructed. One such construction was given in \cite{Mi}. One may then straightforwardly couple $\psi$ to a vector field $A_\mu$ transforming as a fundamental representation covector. 
Further terms, however, in particular kinetic energy terms, apparently cannot be written down in the absence of a metric. Such a gauge theory would automatically be in the `super-strong coupling' limit. 
 
Second, adjoining  translations to form the general or special affine group, one may apply Wigner's argument to classify states of given momentum. Now, however, the little group is $GL(3,R)$, or $SL(3,R)$, so the states are classified according to unitary irreducible representations of these groups. 
There is no notion of particle states in the ordinary sense. The analog of single-particle states 
here are excitations of definite momentum classified according to the infinite-dimensional unitary $GL(3,R)$  representations. 
Amplitudes and correlation functions can be defined in a path integral quantization framework.  
There is, however, no physical interpretations in terms of ordinary particle physics. Such an interpretation emerges only after symmetry breaking 
to the $SO(3,1)$ subgroup. $\overline{SL(4,R)}$ representations generally allow embedding of an infinite 
sum of Lorentz spinors describing a tower of physical spins (see e.g. \cite{KS}). 
Note though that particle fields below the symmetry breaking scale may, in general, be composites of the original fields. 
Also note in this connection that in the broken phase  
the restrictions of  \cite{WW} on the appearance of massless spin-2 particles are evaded 
as in GR; whereas in the unbroken $GL(4,R)$ phase 
there is no invariant notion of helicity, and the argument in \cite{WW} cannot be applied. 
It remains to be seen, of course, whether any consistent $GL(4,R)$-symmetric
theories with spontaneous breaking to the Lorentz group can be constructed. 

\vspace{0.5cm}

The author would like to thank P. Kraus for discussions. This work was partially supported by NSF-PHY-0852438.

\setcounter{equation}{0}
\appendix
\renewcommand{\theequation}{\mbox{\Alph{section}.\arabic{equation}}}
\section{Appendix}
Given some algebra of operators $A, B, \ldots$, define $A\{B\}\equiv [A,B]$, so that  
$A^2\{B\}=[A,[A,B]]$ and so on. One then has the identities
\beq
e^{iA} B e^{-iA} = \sum_{n=0}^\infty {i^n\over n!} \,A^n\{B\}    \label{identI}
\eeq
and 
\beq
e^{-iA(\lambda)} {d\over d\lambda} e^{iA(\lambda)} = i\sum_{n=0}^\infty {(-i)^n\over (n+1)!}
A^n\{{dA\over d\lambda}\}   \, . \label{identII}
\eeq  
Then, from (\ref{identI}) and repeated use of (\ref{CR5}) 
\bea
e^{-{i\over 2}\xi\cdot T} (i {\rm d}x\cdot P)\, e^{{i\over 2}\xi\cdot T} 
&= &i \sum_{n=0}^\infty {i^n\over n!}{1\over 2^n} (-\xi\cdot T)^n\{{\rm d}x\cdot P\} \nonumber \\
& = & i \sum_{n=0}^\infty {i^n\over n!}\, i^n (\xi^n)_{\alpha\beta} {\rm d}x^\beta P^\alpha \nonumber \\
& = & i \left(\exp [ - \xi]\right)_{\alpha\beta} {\rm d}x^\beta P^\alpha  \, .  \label{dgamma1}
\eea
From (\ref{identII}) and repeated use of (\ref{CR2}) - (\ref{CR3}) one gets
\bea
e^{-{i\over 2}\xi\cdot T} {\rm d} e^{{i\over 2}\xi\cdot T}   & = & 
 i\sum_{n=0}^\infty {(-i)^n\over (n+1)!} ({1\over 2}\xi\cdot T)^n\{{1\over 2} {\rm d}\xi\cdot T\}
\nonumber \\
& = & i\left[ \sum_{{\rm even}\atop   n=0}^\infty {(-i)^n\over (n+1)!} {i^n\over 2} 
 \left( \xi^n\{{\rm d}\xi\}\right)_{\alpha\beta} T^{\alpha\beta}  \right. \nonumber \\ 
 &  & \qquad \qquad + \left.
 \sum_{{\rm odd}\atop  n=1}^\infty {(-i)^n\over (n+1)!} {i^n\over 2} 
\left( \xi^n\{{\rm d}\xi\}\right)_{\alpha\beta} J^{\alpha\beta} \right]  \label{dgamma2}
\eea
which, making use of (\ref{identII}) again, can be written as 
\beq
e^{-{i\over 2}\xi\cdot T} {\rm d} e^{{i\over 2}\xi\cdot T}  = {i\over 2} \left[ 
 - {1\over 2} \{e^{\xi} , {\rm d} e^{-\xi}\}_{\alpha\beta}\, T^{\alpha\beta} 
- {1\over 2} [e^{\xi} , {\rm d} e^{-\xi} ]_ {\alpha\beta}\, J^{\alpha\beta}  \right]
\label{dgamma3}
\eeq
with $\{\ ,\ \}$ denoting the anticommutator as usual. 
From (\ref{Gexp1}) and (\ref{dgamma1}), (\ref{dgamma3}) one then has (\ref{Gexp}) and 
(\ref{eform}) - (\ref{spincon1}).

\end{document}